\documentclass[%
 reprint,
 amsmath,amssymb,
 aps,
]{revtex4-2}

\usepackage{graphicx}
\usepackage{xcolor}
\usepackage{dcolumn}
\usepackage{bm}
\usepackage[utf8]{inputenc}
\usepackage[english]{babel}
\usepackage[left=30mm, top=15mm, right=15mm, bottom=15 mm, nohead]{geometry}
\usepackage{subcaption}
\usepackage{graphicx}
\usepackage{wrapfig}
\usepackage{amsmath}

\begin{document}

\preprint{APS/123-QED}

\title{Enhancing the robustness of coupling between a single emitter and a photonic crystal waveguide} 

\author{Alexander Shurinov}
\affiliation{%
 Quantum Technologies Centre, Lomonosov Moscow State University, Russia, Moscow, 119991, Leninskie Gory 1 building 35
}%
\author{Ivan Dyakonov}%
\affiliation{%
 Quantum Technologies Centre, Lomonosov Moscow State University, Russia, Moscow, 119991, Leninskie Gory 1 building 35
}%
 \email{dyakonov@quantum.msu.ru}
\author{Sergei Kulik}
\affiliation{%
 Quantum Technologies Centre, Lomonosov Moscow State University, Russia, Moscow, 119991, Leninskie Gory 1 building 35
}%
\affiliation{%
Laboratory of quantum engineering of light, South Ural State University (SUSU), Russia, Chelyabinsk, 454080, Prospekt Lenina 76
}%
\author{Stanislav Straupe}
\affiliation{%
 Quantum Technologies Centre, Lomonosov Moscow State University, Russia, Moscow, 119991, Leninskie Gory 1 building 35
}%
\affiliation{%
Russian Quantum Center, Russia, Moscow, 121205, Bol'shoy bul'var 30 building 1
}

\date{\today}

\begin{abstract}
We present a heuristic mathematical model of the relation between the geometry of a photonic crystal waveguide and the Purcell enhancement factor at a particular wavelength of interest. We use this model to propose approaches to the design of a photonic crystal waveguide maximizing the Purcell enhancement at a target wavelength. Numerical simulations indicate that the proposed structures exhibit robustness to fabrication defects introduced into photonic crystal geometry.
\end{abstract}

\maketitle


\section{Introduction}\label{sec:introduction}

A planar photonic crystal waveguide (PCW) is a rich system finding applications in diverse areas of optical physics. Among those are slow light \cite{Baba2008}, topological photonics \cite{Wen2022}, chiral photonics \cite{Lodahl2017}, cavity quantum electrodynamics \cite{Vuckovic2003} and many others. An attractive feature of the planar photonic crystal is its flexibility for tailoring dispersion properties of light. In particular, the dispersion curve of a PCW mode inside the crystal bandgap can be engineered to reach extremely high values of group velocity at a target wavelength. This feature allows one to use a PCW as a platform for travelling-wave cavity quantum electrodynamics. A single emitter generating photons at wavelength $\lambda$ matched to the large group velocity range of the PCW mode dispersion curve is strongly coupled to the PCW mode and thus its emission exhibits a significant Purcell enhancement. This effect enabled the development of an on-demand single photon source compatible with planar photonic integrated circuits \cite{Lodahl2020}. Furthermore, the ability to strongly couple an emitter to a cavity mode while still being able to efficiently excite the emitter and read-out photons from the cavity boosted the research in nonlinear light-matter interaction at the single-photon level \cite{Lodahl2015}.

Semiconductor quantum dots (QDs) are the most common type of emitters which can be coupled to a PCW to create a single photon source. Despite being well-studied, QDs with predefined parameters are still notoriously hard to fabricate deterministically. The most widespread Stransky-Krastanov growth process produces QDs with randomly distributed spectral characteristics. The emission wavelength of QDs typically falls in range of a few nanometers around the designed center wavelength. The first derivative $d\omega/dk$ of the PCW dispersion curve gets close to zero in a very narrow wavelength range $\lambda_{0} \pm \Delta \lambda/2$ and efficient Purcell enhancement is not guaranteed for most of the fabricated QDs with emission wavelengths missing the $\Delta\lambda$ region. Furthermore, the fabrication process introduces defects into the PCW structure which affect the dispersion properties of the PCW mode. The workaround for this issue is straightforward -- an array of structures is fabricated and only those which meet particular experimental requirements are selected. Although this method may be satisfactory for research purposes, the lack of reproducibility in the single-photon source fabrication is one of the major bottlenecks in contemporary quantum optical experiments \cite{Pan2021}. At the same time current trends in optical quantum computing demand the development of hybrid integration methods to place single emitters onto a photonic platform of choice \cite{Zwiller2020}. 

In this paper we will address the design approaches which mitigate the effect of fabrication imperfection on the Purcell factor at the source wavelength. We start with developing a heuristic PCW design approach which significantly simplifies the selection of a PCW geometric configuration. The theory behind this approach is based on simple optical phenomena -- interference and diffraction of light scattered inside the PCW membrane and leaking out of the membrane. The derived equations provide clear guidelines how to choose PCW geometric parameters in order to set the maximal Purcell enhancement at the required wavelength and completely eliminate the necessity to evaluate multiple time consuming 3D FDTD simulations. After the description of the heuristic PCW theory we address the problem of PCW robustness to fabrication imperfections. The question of a PCW dispersion curve robustness against fabrication defects has been previously highlighted in the series of works. These include studies of fabrication defects' influence on the quality factors of photonic crystal microcavities \cite{Painter2004, Li2016} and automated design methods to optimize the photonic crystal microcavity structure \cite{Savona2014, Johnson2014}. We focus on the development of a design approach which increases the robustness of the coupling between an emitter and a photonic crystal waveguide mode. We propose two design approaches increasing the robustness of coupling to the fabrication errors and test them using numerical simulations. 

\section{\label{sec:photonic_crystal}Photonic crystal}

\begin{figure*}[!htp]
    \centering
    \includegraphics[width = \textwidth]{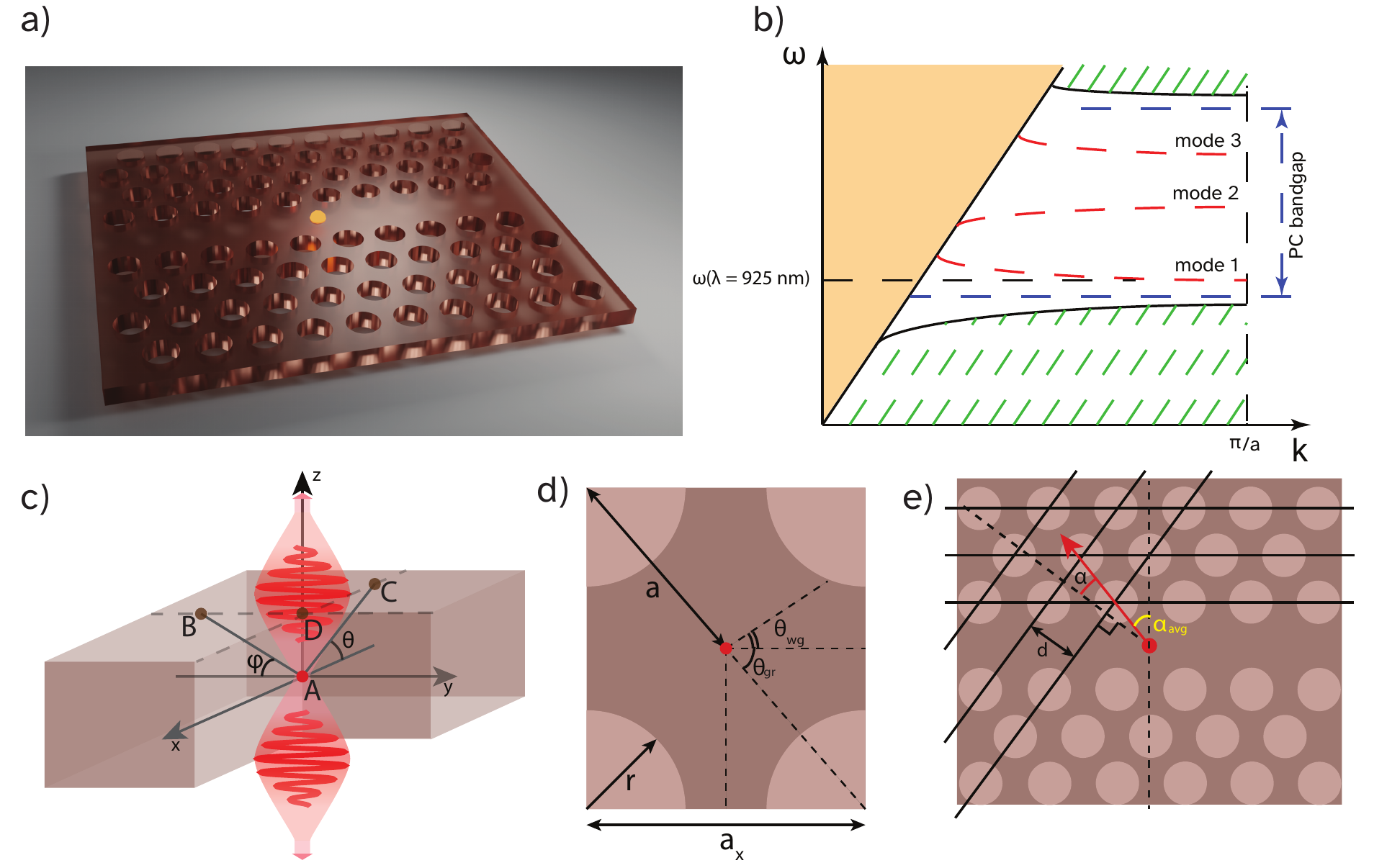}
    \caption{a) An overview of a photonic crystal waveguide structure. b) Typical dispersion curves of PCW eigenmodes inside the photonic bandgap. The example illustrates the existence of 3 modes (red dashed lines). Blue dashed lines indicate the PC bandgap prior to hole removal. Green areas indicate PCW bulk modes and yellow area corresponds to light waves with non-zero wavevector component orthogonal to the PCW membrane surface. Images c), d) and e) illustrate the geometric parameters used throughout the paper.}
    \label{fig:pcw_illustration}
\end{figure*}

A typical two-dimensional photonic crystal is a periodic arrangement of circular holes etched in a thin film of a material with high refractive index. A deleted row of holes forms a photonic crystal waveguide (see illustration in Fig.\ref{fig:pcw_illustration}(a)). A characteristic feature of a PCW is the existence of a frequency range where the group velocity of light decreases significantly. This fact makes a PCW structure an extremely appealing system for mediating interaction between light and an isolated dipole. A PCW effectively serves as a microresonator with small mode volume and high quality factor. These systems were demonstrated to suite the purpose of integration of $A_{3}B_{5}$ quantum dot single photon sources in a planar photonic structure \cite{Lodahl2020}. Quantum dot can be considered as a dipole, which is orientated perpendicular to the waveguide axis in the PC plane. A PCW microresonator forms an open cavity which can be smoothly interfaced with other integrated photonic waveguides. In this paper we study methods to increase robustness of PCW features to fabrication defects.

We focus our attention on a PCW created by deleting a row of air holes from a 2D triangular array. The host material is chosen to be gallium arsenide (GaAs) because the target application is a planar semiconductor quantum dot single photon source. We start with the description of PCW characteristics and development of its heuristic model. Manga Rao and Hughes \cite{Manga_Rao} derived an expression for a Purcell factor $F_{p}$ in terms of PCW parameters:

\begin{equation}\label{eq:purcell_factor_pcw}
    F_{p} = \frac{3\pi c^{3}a}{V_{eff}\omega_{d}^{2} \epsilon^{3/2}v_{g}},
\end{equation}
where $a$ is the distance between air holes (if the lattice is triangular $a=a_x$, since we are focused on that type of lattice from now on we will write $a$ instead $a_x$, but for another lattice angle $a_x$ is the only correct option), $V_{eff}$ is the effective mode volume and $v_{g}$ is the group velocity at the resonant frequency of the dipole $\omega_{d}$. The formula indicates that the largest $F_{p}$ is achieved when the wavepacket group velocity reaches zero. Thus the design of a PCW efficiently coupled with a single emitter resonant at $\omega_{d}$ is equivalent to engineering a PCW dispersion law to meet the requirement $d\omega/dk(\omega_{d})=0$. Numerical methods for calculation of the dispersion structure of a PCW are well-known and straightforward \cite{Lodahl} and can be easily applied to a PCW with a defined geometry. However, there exists no recipe of how to estimate geometrical parameters of a PCW exhibiting high Purcell factor at a wavelength of interest. We devise heuristic expressions linking the target wavelength and the parameters of a hexagonal PCW which stem upon simple optical effects taking place inside a photonic crystal. Based on these results we introduce methods to increase the robustness of a PCW structure to fabrication defects.

\section{\label{sec:purcell_peak_fit}A PCW Purcell factor heuristics}

\begin{figure}
    \centering
    \includegraphics[scale=0.4]{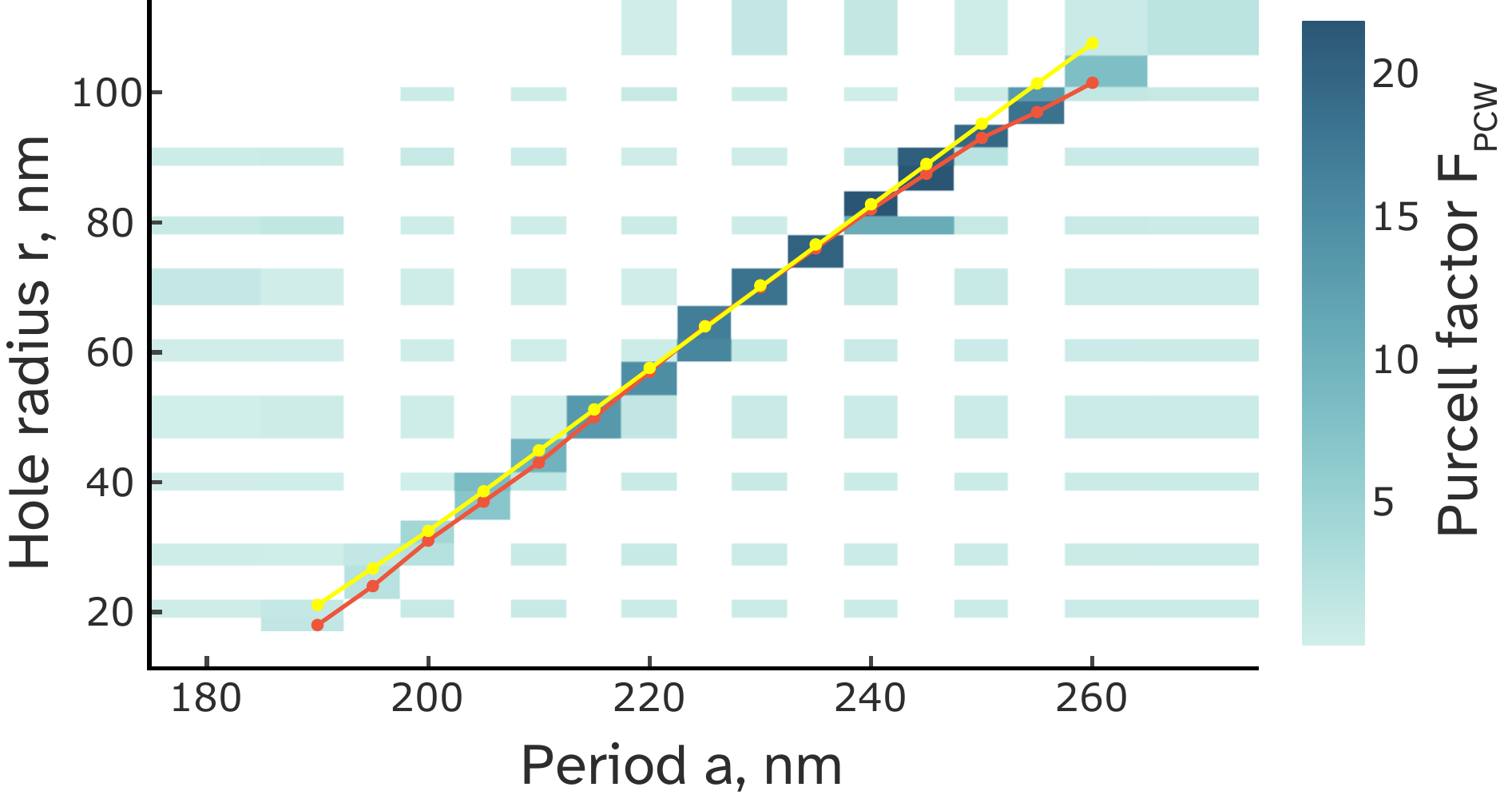}
    \caption{The set of configurations in $(a,r)$ parameter space corresponding to a peak $F_{PCW}$ at 925 nm. The 2D heatmap represents the $F_{PCW}$ values experienced by the emitter in a PCW with given $a$ and $r$. The $a$ and $r$ parameter values are unevenly distributed for time-saving purposes. We added extra points in the maximal $F_{PCW}$ region. The white tiles represent the points which were not computed. The red curve connects the $F_{PCW}$ maximal values calculated numerically using 3D FDTD simulation. The yellow curve contains values estimated by the proposed heuristic theoretical approach.}
    \label{fig:purcell_peak}
\end{figure}

Figure~\ref{fig:pcw_illustration}(b) illustrates a typical dispersion structure of a PCW. The geometrical parameters for this example are as follows: PC hole pattern angle $\theta_{gr}=60^{\circ}$, period $a=0.238$~$\mu$m, hole radius $r=0.08$~$\mu$m and membrane thickness $h=0.16$~$\mu$m. These values were chosen to put the Purcell factor $F_{PCW}$ peak at $925$~nm. A natural question arises whether this configuration is unique. It turns out that the answer is negative. We performed an extensive numerical analysis of the Purcell enhancement happening in different PCW configurations, results are presented in Fig.~\ref{fig:purcell_peak}.The 3D FDTD simulation was carried out in Lumerical FDTD package, the details of the simulation are specified in Appendix~\ref{app:simulation}. We observed a continuous set of configurations of a triangular PCW with the same lattice angle corresponding to a peak value of $F_{PCW}$ at a target wavelength. The red line in Fig.~\ref{fig:purcell_peak} illustrates the numerically computed set of $(a,r)$ configurations corresponding to the most efficient coupling of a PCW mode to dipole radiation at $925$~nm. The curve in $(a,r)$ space closely follows the function $a=c_{1}+c_{2}r$, where the coefficients $c_{1}$ and $c_{2}$ are weakly dependent on $a$ and $r$. The $a(r)$ dependence is finely approximated by a linear function in the region where the $F_{PCW}$ reaches its highest levels. The yellow curve represents the values of $a$ and $r$ corresponding to $F_{PCW}$ peak which are provided by the proposed theoretical description.

In the following subsections we provide a heuristic theoretical description for the origins of such dependence and the values of $c_{1}$ and $c_{2}$ coefficients.

\subsection{The slope coefficient $c_{2}$}
The Purcell factor $F_{PCW}$ defines the probability
\begin{equation}\label{eq:pcw_coupling_probability}
    \beta =F_{PCW}/(1+F_{PCW})
\end{equation}
of emitting a photon into a PCW mode. The existence of a PCW mode is a purely interferometric effect hence the probability $p$ should be related to the geometry of a photonic crystal. We expect to derive the connection between the geometric parameters which correspond to a configuration of a PCW structure reaching maximal Purcell factor for the required wavelength. We roughly split the emitter radiation into three categories: light exiting the PCW plane, light propagating inside the PCW structure, and light coupled to the PCW mode. For the light exiting the PCW plane we define the notion of a vertical Fabry-Perot resonator and the corresponding Purcell factor $F_{FP}$, which is used to evaluate a portion of light leaking from the PCW membrane. Then the fraction of light emitted in the PCW itself from the total amount of radiation which remains inside the crystal can be estimated using the effective angle $\theta_{wg}$ (see Fig.~\ref{fig:pcw_illustration}(d)). Under such assumptions we can derive the following equation:
\begin{equation}\label{eq:detailed_main_relation}
    \left(1-\frac{F_{FP}}{1+F_{FP}}\right)\frac{\int_{\pi/2-\theta_{wg}}^{\pi/2+\theta_{wg}}\sin^3\theta{d}\theta}{\int_0^\pi\sin^3\theta{d}\theta}=\frac{F_{PCW}(a,r)}{1+F_{PCW}(a,r)},
\end{equation} 
where the first term on the left-hand side of the equation denotes the probability of the photon to stay inside the crystal and the second term denotes the fraction of the photons emitted into the waveguide mode. Here we assumed that if the total internal reflection angle is relatively small (approximately $16^{\circ}$ in case of GaAs to air transition), then the majority of the photons which leak out of the crystal can be attributed to the emission into the Fabry-Perot resonator mode.  The right-hand side accounts for the probability of emitting the photon exactly into the waveguide mode using the $F_{PCW}$ value. Here we have three variables, which we need to calculate: $F_{FP}$, $F_{PCW}$ and $\theta_{wg}$. We calculate values $F_{FP}$ and $F_{PCW}$ and use equation \ref{eq:detailed_main_relation} to determine the value of $\theta_{wg}$.

The $\theta_{wg}$ is different for each individual crystal configuration. The expression connecting geometrical parameters of a crystal to the $\theta_{wg}$ value is defined by the crystal configuration. For the $60^{\circ}$ triangular hole pattern the $\theta_{wg}$ definition is illustrated in Fig.~\ref{fig:pcw_illustration}(d).
For a triangular lattice, the angle $\theta_{wg}$ is implicitly related to the period $a$ and the hole radius $r$ (see Appendix A):
\begin{equation}\label{eq:a_to_r_ratio}
    \frac{a}{r}=\frac{\cos(\pi/4-\theta_{wg}/2)-\tan\theta_{wg}\sin(\pi/4-\theta_{wg}/2)}{\tan\theta_{gr}/2-\tan\theta_{wg}/2},
\end{equation}
where $\theta_{gr}$ is the lattice angle. Once we have estimated $F_{FP}, F_{PCW}$, and $\theta_{wg}$ values we can substitute them into equation \ref{eq:a_to_r_ratio} and evaluate the $c_{2} = a/r$ coefficient.

To calculate $F_{FP}$ we note, that light emitted by a dipole at frequency falling into the photonic crystal bandgap exhibits propagation through a low-quality Fabry-Perot resonator between the top and the bottom surfaces of the PCW membrane. Resonance frequencies $\omega_{m}=(\pi c/nh)m=\frac{\pi\cdot{c}}{nh}=w_{1}$ and the linewidth $dw=c(1-R)/(n\sqrt{R}h)$ for m=1 of this Fabry-Perot resonator are expressed using the membrane thickness $h$ and the refractive index of the material $n$. Fresnel law yields the reflection coefficient $R=(n-1)^2/(n+1)^2$ for light incident normally to the membrane surfaces. A fraction of light emitted by the dipole into the vertical Fabry-Perot mode can be estimated using a Purcell factor 

\begin{equation}\label{eq:purcell_factor_vertical_fp}
    F_{FP} = \frac{3Q (\lambda/n)^3}{4\pi^{2}V_{eff}} \cdot \frac{d\omega^{2}}{4(\omega - \omega_{1})^{2}+d\omega^{2}},
\end{equation}
 From this equations we obtain the quality factor $Q=\frac{w_1}{dw}$. The effective mode volume is given by $V_{eff}=\frac{\int_{V}{\epsilon{E^2}}dV}{\underset{V}{\max}(\epsilon{E^2})}$ \cite{Hughes2004}, where the integration is carried out over a single unit cell for periodical structures (for example PCWs) and over the whole possible volume for non-periodical structures (the period equals to infinity). For such configuration the effective mode volume can be estimated as (see Appendix B) $V_{eff}=\frac{2}{3}h^3$. The $F_{FP}$ of the vertical mode resonator then equals to $F_{FP}\approx 0.64$ if we set $\lambda=925$~nm and $n_{GaAs}(\lambda = 925\,\mathrm{nm})=3.46$.

Next we need to express $F_{PCW}(a,r)$. The effective volume of the PCW mode $V_{eff}$ from Eq.~\ref{eq:purcell_factor_pcw} is the last term which is not yet related to $a$ and $r$. We consider a unit cell as a rectangle with 4 quarter circles (see Fig.~\ref{fig:pcw_illustration}(d)). The effective mode volume for the light modes with the free-space dispersion relation are defined by
\begin{equation}\label{eq:free-space_mode_volume}
    V_{eff}=\frac{\int_V I dV}{\underset{V}{\max I}}=\frac{\langle I\rangle V}{I_{0}},
\end{equation}
where $I$ denotes the radiation intensity inside the unit cell. We consider a PCW dispersion curve with $d\omega/dk(\omega_{d})=0$ and focus on the system behaviour at a frequency $\omega$ slightly smaller than $\omega_{d}$. We assume that the mode volume of light states at the frequencies $\omega < \omega_{d}$ is roughly the same as at the resonance frequency  $\omega_{d}$, because the slight change of the light frequency shouldn't drastically affect the mode volume. We also assumed that the emitter is preferentially coupled to the primary mode (mode 1 in Fig.~\ref{fig:pcw_illustration}b). Light at frequency $\omega$ can only populate the states satisfying the free space dispersion relation $\omega = ck/n(\omega)$. The dipole intensity is proportional to $\frac{1}{r^2}$ and the intensity at the 'entrance' of the unit  cell is $I_0$. We can find the function $I(x)=\frac{const}{(x+1)^2}$, where $x$ is the dimensionless coordinate along the axis of the waveguide, $const$ equals numerically to $I_0$, and $x=0$ at the 'entrance' of the unit cell. Using $I(x)$ we can easily calculate the average intensity in the unit cell

\begin{equation}\label{eq:average_intens_unit_cell}
    \langle I(x)\rangle =\frac{1}{a}\int_{0}^{a}\frac{const}{(x+1)^2}dx=\frac{const}{(\langle x \rangle +1)^{2}},
\end{equation}
where $a$ is the lattice period along the propagation axis. This result implies, that we need to consider $\langle I \rangle$ as the intensity at the $\langle x \rangle$ coordinate taking into account diffraction effects at the unit cell 'entrance'. The estimated number of Fresnel zones (we assume, that the unit cell is located far from the dipole, so the light propagates almost parallel to the waveguide axis)
\begin{equation}\label{eq:fresnel_zones}
    m=\frac{(a\tan\theta_{gr}-2r)^2}{4\lambda<x>}\approx1.2 ,
\end{equation}
which means that the Fresnel approximation is eligible when accounting for the diffraction effects. Since the unit cell entrance cross-section is rectangular, we calculate the parameters of the Cornu spiral as 
\begin{equation}\label{eq:cornu_parameters}
\begin{array}{c}
u=\sqrt{2m},\\
c(u)=\int_{0}^u \cos(\frac{\pi}{2}\tau^2)d\tau,\\
s(u)=\int_{0}^u \sin(\frac{\pi}{2}\tau^2)d\tau.
\end{array}
\end{equation}
Coefficients $c(u)$ and $s(u)$ allow to evaluate the required ratio
\begin{equation}\label{eq:intensity_ratio}
    \frac{\langle I\rangle}{I_0}=\frac{c(w)^2+s(w)^2}{c(\infty)^2+s(\infty)^2}.
\end{equation}
Then the effective mode volume is equal to $V_{eff}=\frac{\langle I\rangle}{I_0}V$, where 
$V$ is the geometrical volume of the unit cell. 

Now we have all the ingredients to specify the explicit relation $a(r)$. The $F_{PCW}$, and $F_{FP}$ are all expressed as functions of $a$ and $r$ and after substitution of each one to the equation (\ref{eq:detailed_main_relation}) we get $\theta_{wg}$(a,r) and equation (\ref{eq:a_to_r_ratio}) gives $\frac{a}{r}=c_{2}$. The explicit formulas and a numerical calculation algorithm are provided in Appendix B.  

\subsection{The constant coefficient $c_1$}

When the hole radius $r$ is close to zero, the PC becomes analogous to a Bragg grating (see Fig.~\ref{fig:pcw_illustration}). When the $F_{PCW}$ reaches its' peak, the crystal blocks all the photons, which are emitted outside the waveguide direction. The Bragg equations define destructive interference criteria for the photons, which are not emitted to the PCW mode. We will only take into account two types of reflective surfaces: the parallel and the sloped as shown in Fig.~\ref{fig:pcw_illustration}(e).These two sets of surfaces share similar properties: the distance between the circles along these surfaces is minimal over all other possible surfaces and equals to the crystal parameter $a$. We do not take into account the interference effects happening on every other possible set of surfaces due to the increasing distance between the circles and thus the necessity to take diffraction effects into consideration. For the sloped surfaces the Bragg condition is 
\begin{equation}\label{eq:sloped_bragg_condition}
    \begin{array}{c}
         2dn\cos(\alpha)=\lambda,\\
         \alpha=\theta_{gr}-\alpha_{\mathrm{avg}},\\
         d=a_{slp}\sin(\theta_{gr}),
    \end{array}
\end{equation}
where  $a_{slp}$ is the lattice period, found using the Bragg condition for sloped surfaces, and $\alpha_{\mathrm{avg}}$ is the average angle of emission, which can be determined using the following formula 
\begin{equation}\label{eq:average_angle}
    \alpha_{\mathrm{avg}}=\frac{2}{\pi}\int_0^{\pi/2}\sin^3\theta\cos\theta{d}\theta.
\end{equation}
To obtain this formula we take into account that we need to calculate the angle corresponding to the average direction of power emission in any quarter surface. The term $\sin^2\theta$ describes the dipole radiation pattern. Another $\sin\theta$ term arises from the transition to a spherical coordinate system. Lastly the $\cos\theta$ term implies, that we are interested in a projection on the $y$-axis, because we look for the destructive interference condition for light waves propagating perpendicular to the waveguide axis. 

The Bragg condition for the parallel surfaces is
\begin{equation}\label{eq:parallel_bragg}
    2\frac{a_{par}\cdot\tan{\theta_{gr}}}{2}n\cos\alpha_{\mathrm{avg}}=\lambda,
\end{equation}
where $a_{par}$ is the lattice period found using the Bragg condition for the parallel surfaces. Finally we need to estimate the fractions of radiation which preferentially interfere in the sloped and the parallel Bragg gratings. These fractions could be obtained using the dipole radiation pattern
\begin{equation}\label{eq:sloped_and_parallel_fractions}
\begin{array}{c}
    \eta_{\mathrm{slp}}=\frac{\int_{\pi/2-\theta_{gr}}^{\pi/2}\sin^3\theta{d}\theta}{\int_{0}^{\pi/2}\sin^3\theta{d}\theta},\\
    \eta_{\mathrm{par}}=\frac{\int_{0}^{\pi/2-\theta_{gr}}\sin^3\theta{d}\theta}{\int_{0}^{\pi/2}\sin^3\theta{d}\theta}.\\
 \end{array}   
\end{equation}
Taking these fractions into account, we can now express the total constant coefficient through $a$ and $r$:
\begin{equation}\label{eq:constant_coefficient_value}
    c_{1}=\eta_{par}a_{par}+\eta_{slp}a_{slp}.
\end{equation}

\label{sec:robustness}
\section{Improvement of the PCW robustness to fabrication imperfections using composite PCW structures}

\begin{figure}[t!]
    \centering
    \includegraphics[width = \linewidth]{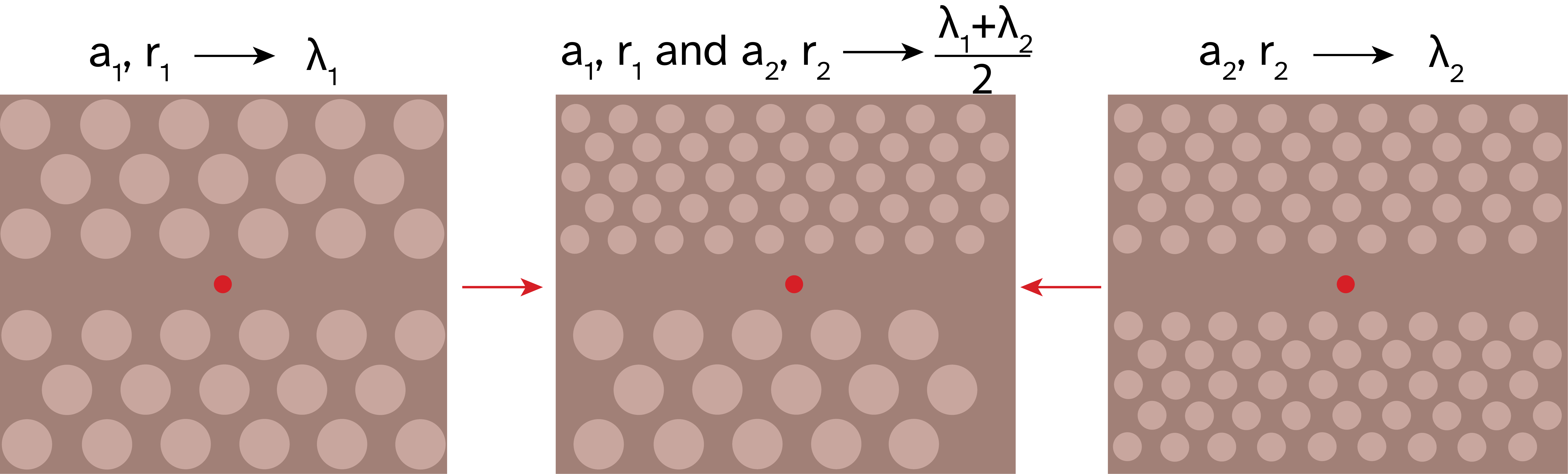}
    \caption{Compound PCW composed of two half-PCWs, designed to exhibit peak value of $F_{PCW}$ at $\lambda_{1}$ and $\lambda_{2}$, which satisfy the condition $\lambda_{1}+\lambda_{2}=2\lambda_{0}$. }
    \label{fig:compound_crystal}
\end{figure}

The first method is based on Eq.~\ref{eq:a_to_r_ratio}. We assume, that $\theta_{wg}$ remains the same for all quarters of the PCW unit cell. However, it turns out that this angle shouldn't necessarily be realised by the same lattice period $a$ and radius $r$ in each quarter. We state that if two PCWs with different lattice periods $a_{1}, a_{2}$ and hole radii $r_{1}, r_{2}$, respectively, reveal the peak of $F_{PCW}$ at the same wavelength, then the compound crystal will also reveal the peak at the same wavelength (see fig. \ref{fig:compound_crystal}).

We suggest that the PCW in the center of Figure~\ref{fig:compound_crystal} is more robust to the fabrication imperfections than the one on the right, and the one on the left. If we introduce slight random deviations to the values of radius $r$, the $F_{PCW}$ spectral curve shifts away from the target wavelength and the coupling of the dipole radiation to the PCW mode decreases substantially. We do not take into account random deviations of the PCW period $a$ because its value is several times larger than the deviation due to manufacturing imprecision introduced in state-of-the-art fabrication lines \cite{Fan2010}. It's also worth noting that a systematic bias of $a$ and $r$ is easily accounted for by our theoretical description. If the bias in each of the parameters can be determined experimentally, the other parameter can be adjusted accordingly using the formulas for $c_{1}$ and $c_{2}$ coefficients. The idea behind increased robustness to fabrication errors relies on simple reasoning. The $\beta$ factor value gets higher than $0.9$ even at moderate $F_{PCW} \approx 10$ or greater, which can already be considered as good coupling of the emitter radiation to the PCW mode. The width of the $F_{PCW}$ spectral dependence is extremely narrow due to the mode dispersion curve (see fig \ref{fig:pcw_illustration}) of an ideal PCW. Our goal is to 'spoil' the PCW structure in order to make the $F_{PCW}$ spectrally wider at the expense of the $F_{PCW}$ peak value becoming lower but still sufficient for good coupling. 
\begin{figure}[t!]
    \centering
    \includegraphics[width=\linewidth]{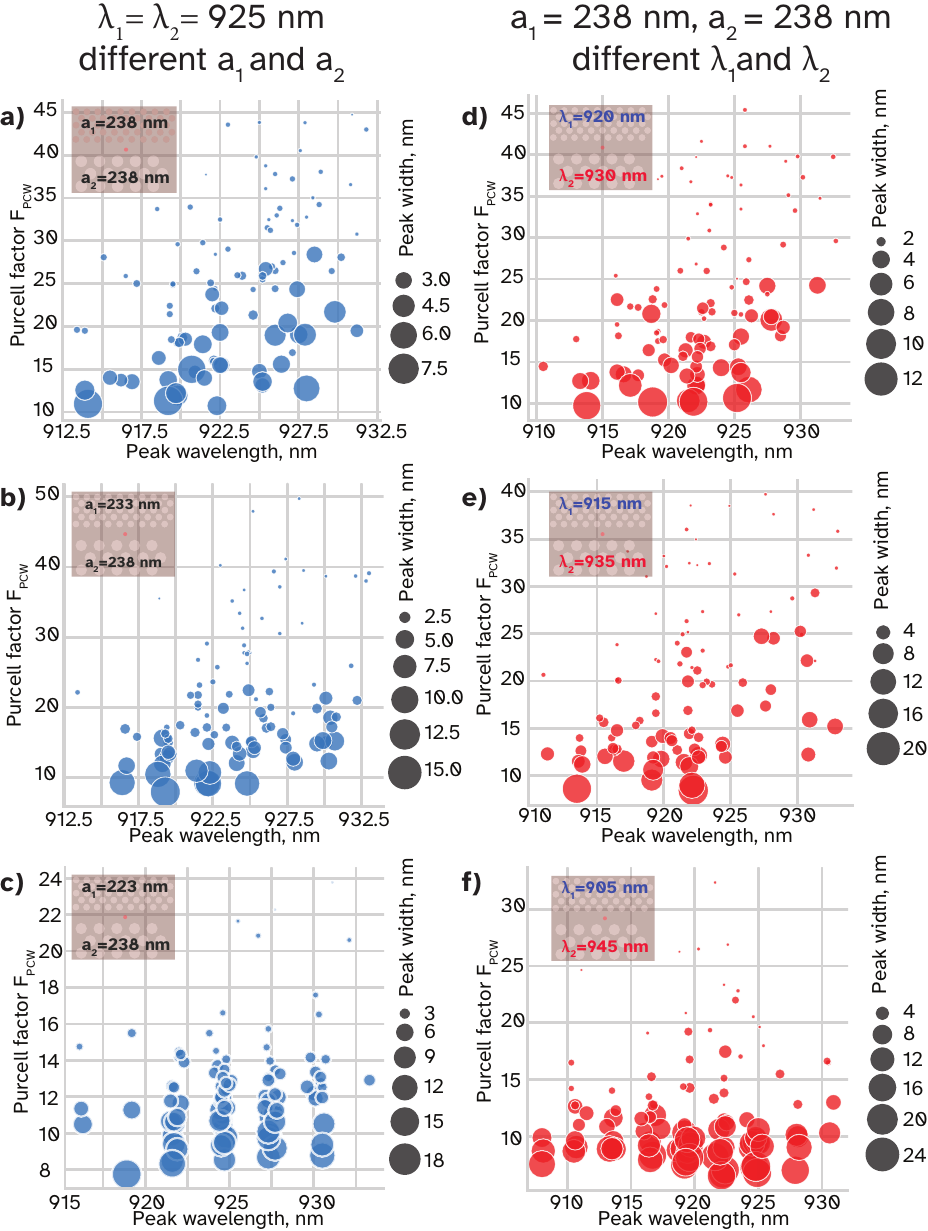}
    \caption{The graphs shows summarized results of the simulations of the compound crystals $F_{PCW}$ spectral curves. Each panel depicts the $F_{PCW}$ peak values and FWHMs for each of the 100 curves computed for compound PCWs with added randomized hole radius error. The left column shows the results for compound structures composed of two halves of PCW optimized for identical peak $F_{PCW}$ wavelength $\lambda_{0}=925$~nm. The right column shows the results for compound structures comprised of two halves of PCWs optimized for different $\lambda_{1}$ and $\lambda_{2}$ satisfying $\lambda_{1}+\lambda_{2} = 2\lambda_{0}$.The insets in each figure illustrate the used set of variable parameters.}
    \label{fig:compound_crystal_simulation_results}
\end{figure}

We prove our point by performing numerical simulations using the Lumerical software package. The model describes a composite PCW structure comprised of two half-crystals with different periods and hole radii. Both halves correspond to the PCWs delivering optimal $\beta$ at the same wavelength $\lambda = 925$~nm. We add random additive $\delta r_{i}$ sampled from the range of $[-10,10]$~nm to the radius each hole in the numerical PCW model and perform 100 simulation runs. Figure~\ref{fig:compound_crystal_simulation_results} illustrates the results of the simulation. The best combination corresponds to the compound crystal with periods $a_{1}=233$~nm and $a_{2}=238$~nm respectively. The geometry of both parts of the compound crystal was estimated using our model and each one reveals the highest value of $F_{PCW}$ at $925$~nm. This compound crystal configuration demonstrates better performance (Fig. \ref{fig:compound_crystal_simulation_results}, left column, central panel) with simulated fabrication defects compared to the standard PCW with identical halves (Fig. \ref{fig:compound_crystal_simulation_results} top row, left panel). To quantify the performance we introduce an average Purcell enhancement factor $\tilde{F}_{PCW}$ and a probability of Purcell enhancement $p$. The target wavelength lies within the width at half maximum of Purcell enhancement curve in $p\times 100$ instances of simulations, where $p$ is the probability of Purcell enhancement. The average $\tilde{F}_{PCW}$ is a mean value of $F_{PCW}$ in these instances. In the case of the compound crystal with randomized radii the probability of Purcell enhancement at $925$~nm equals to $p=0.35$ and the average value $\tilde{F}_{PCW} = 8.69$, whereas for the standard crystal the values are 24 out of 100 and $\tilde{F}_{PCW}=7.62$ respectively, confirming the robustness of the compound PCW in comparison with the non-compound one.

Another option is to consider a compound PCW made of two parts designed to exhibit maximal $F_{PCW}$ value at different wavelengths $\lambda_{1}$ and $\lambda_{2}$. Here the question arises: at which wavelength the maximum $F_{PCW}$ of a compund crystal will be observed? The answer turns out to be simple: If the parts of the PCW reveal the peak $F_{PCW}$ value at wavelengths $\lambda_1$ and $\lambda_2$ respectively, then the compound PCW reveals the peak at $\frac{\lambda_1+\lambda_2}{2}$. One can easily prove this statement by taking into account that there are two different reflective surface systems (under an approximation of a small hole radius $r$, see Fig.~\ref{fig:pcw_illustration}e), and in order to prevent light from propagation in a direction perpendicular to the waveguide axis, the effective distance between each surface must be equal to $\lambda/2$,  meaning that the total effective distance between the surfaces on both sides from the emitter should be $\frac{\lambda_1+\lambda_2}{2}$.

\begin{table*}
\begin{tabular}{c|c|c|c|c}
 Configuration & $\tilde{F}_{PCW}$ @ 925 nm & p & mean FWHM, nm & mean $F_{PCW}$ \\ 
 \hline
 $\lambda_{1}=\lambda_{2}=925$~nm and $a_{1}=a_{2}=238$~nm & 7.6 & 0.24 & 3.0 & 24.9 \\
 $\lambda_{1}=\lambda_{2}=925$~nm and $a_{1}=233$~nm, $a_{2}=238$~nm & 8.6 & 0.35 & 4.0 & 21.4 \\
 $\lambda_{1}=\lambda_{2}=925$~nm and $a_{1}=223$~nm, $a_{2}=238$~nm & 7.0 & 0.57 & 7.4 & 10.2 \\
 $\lambda_{1}=920$~nm, $\lambda_{2}=930$~nm and $a_{1}=a_{2}=238$~nm & 6.9 & 0.26 & 3.5 & 23.8 \\
 $\lambda_{1}=915$~nm, $\lambda_{2}=935$~nm and $a_{1}=a_{2}=238$~nm & 5.4 & 0.18 & 4.4 & 20.6 \\
 $\lambda_{1}=905$~nm, $\lambda_{2}=945$~nm and $a_{1}=a_{2}=238$~nm & 4.6 & 0.44 & 11.1 & 12.5 \\
 \end{tabular}
 \caption{The summary of the combined crystal performance according to numerical simulations. The mean FWHM and mean $F_{PCW}$ values indicate the Purcell factor FWHM and maximal value averaged over 100 simulation runs.}
 \label{tab:result_summary}
\end{table*}

The numerical results for this composition method are also obtained using 100 FDTD simulations similar to those used previously. The results are shown in the right column of Fig.~\ref{fig:compound_crystal_simulation_results}. The conclusion for this case is the following: the greater is the difference between the wavelengths of both parts of a compound PCW, the less is the maximal value and the greater is the FWHM of the $F_{PCW}$ spectral curve. The periods for both parts can be the same or different, but the radius value must be set according to our model, so that each part of the PCW shows a peak at the required wavelength.

Results of the simulation of the Purcell factor spectral curve in the combined crystals with randomized radius error $\delta r \in [-10,10]$~nm are summarized in Table \ref{tab:result_summary}. We observe the tendency of the probability $p$ to grow when the two halves of the PCW are designed with largely different parameters.

\section{Discussion and conclusion}\label{sec:discussion}
We have established a mathematical connection between geometrical parameters of a PCW structure and a Purcell enhancement factor at the specific wavelength. Compound PCW structures which are predicted to provide maximal enhancement at the target wavelength exhibit stronger robustness to the random hole radius deviation compared to standard PCW structures. We attribute the observed effect to the broken symmetry of the crystal. The improvement manifests itself in higher probability of a dipole emitter to be efficiently coupled to a PCW mode and the higher average Purcell enhancement factor $\tilde{F}_{PCW}$. The $\tilde{F}_{PCW}$ values are around $\approx 10$ which yields the $\beta$ factor to be $\approx 91\%$. Although such coupling efficiency values cannot be considered as satisfactory for the most demanding applications like fault-tolerant linear optical quantum computing \cite{Rudolph2021, Jeong2022}, they can nevertheless enable near-term experiments with multiple single-photon sources on an integrated platform.

We would like to draw the readers' attention to a few interesting features which were uncovered in course of the simulations with randomized radii deviations. The configurations designed for identical target wavelengths $\lambda_{1}=\lambda_{2}=925$~nm with unequal periods $a_{1} \neq a_{2}$ (see Fig.~\ref{fig:compound_crystal_simulation_results})~b and c) both show clustering of points. The configurations in Fig.~\ref{fig:compound_crystal_simulation_results}c and Fig.~\ref{fig:compound_crystal_simulation_results}f have the most pronounced clustering along four vertical lines. We were unable to explain the origin of this behaviour, but we speculate that this effect might be related to the emergence of a topologically protected mode inside the PCW bandgap \cite{Proctor2020}. Another peculiar observation is the tendency to cluster along the vertical lines separated by an almost equal distance $\Delta \lambda$. This means that a discrete set of wavelengths exhibits a highly robust Purcell enhancement in the presence of the randomized hole radius error.

Our results provide a clear understanding of the PCW parameter interplay and thus they significantly simplify the initial structure design procedure. They can also serve for augmenting sophisticated automated optimization design routines by narrowing down the parameter space or serving as a quick sanity check avoiding the necessity to run a 3D FDTD simulation task. Our heuristic model describes a triangular PCW structure only, but we believe that similar reasoning and mathematical analysis apply to any other photonic crystal layout. 

\section{Acknowledgements}

A. S., I. D. and S. S. acknowledge support by Rosatom in the framework of the Roadmap for Quantum computing (Contract No. 868-1.3-15/15-2021 dated October 5, 2021 and Contract No.P2154 dated November 24, 2021). S. K. is supported by the Ministry of Science and Higher Education of the Russian Federation on the basis of the FSAEIHE SUSU (NRU) (Agreement No. 075-15-2022-1116).


\bibliography{manuscript}

\appendix

\section{The $\frac{a}{r}$ formula}
The idea underlying the derivation of the formula for $\theta_{wg}$ is the following: $\theta_{wg}$ is an angle between the waveguide's axis and a beam which is reflected without obtaining a component parallel to the waveguide's axis (see Fig.~\ref{fig:pcw_illustration}d). Then we can use simple geometric relations to get 
\begin{equation}
    \tan\theta_{wg}=\frac{AB}{BC}=\frac{a\tan\theta_{gr}/2-r\cos\alpha}{a/2-r\sin\alpha}
\end{equation}
and $\alpha=\pi/4-\theta_{wg}/2$. Once we combine these relations we get the equation
\begin{equation}
    \tan\theta_{wg}=\frac{a\tan\theta_{gr}/2-r\cos(\pi/4-\theta_{wg}/2)}{a/2-r\sin(\pi/4-\theta_{wg}/2)},
\end{equation}
which yields the formula in the form of eq. \ref{eq:a_to_r_ratio}:
\begin{equation}
 \frac{a}{r}=\frac{\tan\theta_{wg}\sin(\pi/4-\theta_{wg}/2)-\cos(\pi/4-\theta_{wg}/2)}{\tan\theta_{wg}/2-\tan\theta_{gr}/2}.
\end{equation}
 
\section{Estimated value of the effective mode volume for a flat surface Fabry-Perot resonator}
We use a simple relation to estimate an effective mode volume $V_{eff}=h\cdot{S_{eff}}$ of the Fabry-Perot resonator, comprised of the top and bottom surfaces of the PCW membrane. We take into account the angular structure of the dipole radiation which has the form of the product of two functions $f(\theta)f(\phi)$. Based on that we can estimate the effective mode area $S_{eff}$ as a product of effective mode lengths in both directions $\theta$ and $\phi$ (see Fig. \ref{fig:pcw_illustration}c). The effective mode length for each direction can be derived as 
\begin{equation}\label{eq:effective_length}
 l_{eff}=\frac{\int_{L} I dl}{I_{max}}.    
\end{equation}
The intensity value is inversely proportional to the distance between the source and the observation point and hence we have $I_{max}\sim\frac{1}{(h/2)^2}$ and $I\sim\frac{1}{AC^2}$. We use geometric relations $AC=\frac{h/2}{\sin\theta}$, $DC=h/2\cot\theta$,  $dl=d(DC)=\frac{h/2}{\sin^2\theta}d\theta$ and substitute them into equation (\ref{eq:effective_length}). We also need to account for the dipole radiation angular dependence $\sin^2\theta$ and the term $\sin\theta$ corresponding to projection on the $z$ axis, because we are only considering light states propagating perpendicular to the PCW membrane. The dependence on the azimuthal angle $\phi$ accumulates only a single $\sin\phi$ term due to the absence of dependency on $\phi$ in the dipole emission pattern. The final integrals for both effective lengths are 
\begin{equation}\label{eq:effective_length_integral}
\begin{array}{c}
    l_{\theta}\sim\frac{1}{I_{max}}\int_{0}^{\pi}{\frac{\sin^3(\theta)}{h/2}d\theta},\\
    l_{\phi}\sim\frac{1}{I_{max}}\int_{0}^{\pi}{\frac{sin\phi d\phi}{h/2}}.
    \end{array}
\end{equation}

The integration yields effective length values $l_{\theta}\sim\frac{2}{3}h$ and $\l_{\phi}=h$. The total effective mode volume then equals to $V_{eff}=h\cdot\frac{2}{3}h\cdot{h}=\frac{2}{3}h^3$.\\

\section{The calculation algorithm}
One can use the following procedure in order to calculate the value of the hole radius $r$ given the parameters $h, n, \lambda_{d}, a$ :\\
\begin{enumerate}
    \item Set an initial guess $r_{0}$ for the radius value for the given $a$, it should be relatively close to the exact value;
    \item Calculate the $F_{FP}$ using equation (\ref{eq:purcell_factor_vertical_fp});
    \item Calculate the $F_{PCW}$ using equations (\ref{eq:free-space_mode_volume})-(\ref{eq:intensity_ratio}) and use $r_{0}$ value defined at step 1. After calculating the PCW mode volume value substitute it into equation (\ref{eq:purcell_factor_pcw});
    \item Calculate the slope coefficient $c_{2}$ for the given $a$ using equation (\ref{eq:a_to_r_ratio});
    \item Calculate the $c_{1}$ coefficient using equations (\ref{eq:sloped_bragg_condition})-(\ref{eq:constant_coefficient_value});
    \item The exact radius for the model can now be estimated as $r=\frac{a-c_{1}}{c_{2}}$;
    \item Repeat all the steps using $r$ as the new initial value. This should be done until the initial value of radius set at step 1 equals to the exact value, which was obtained at step 6.
\end{enumerate}

 Such procedure is necessary due to the explicit dependence $r(a)$, and it is the easiest way to obtain the result.

 \section{The FDTD simulation details}\label{app:simulation}
 
  \begin{figure}[t!]
 	\centering
 	\includegraphics[width=\linewidth]{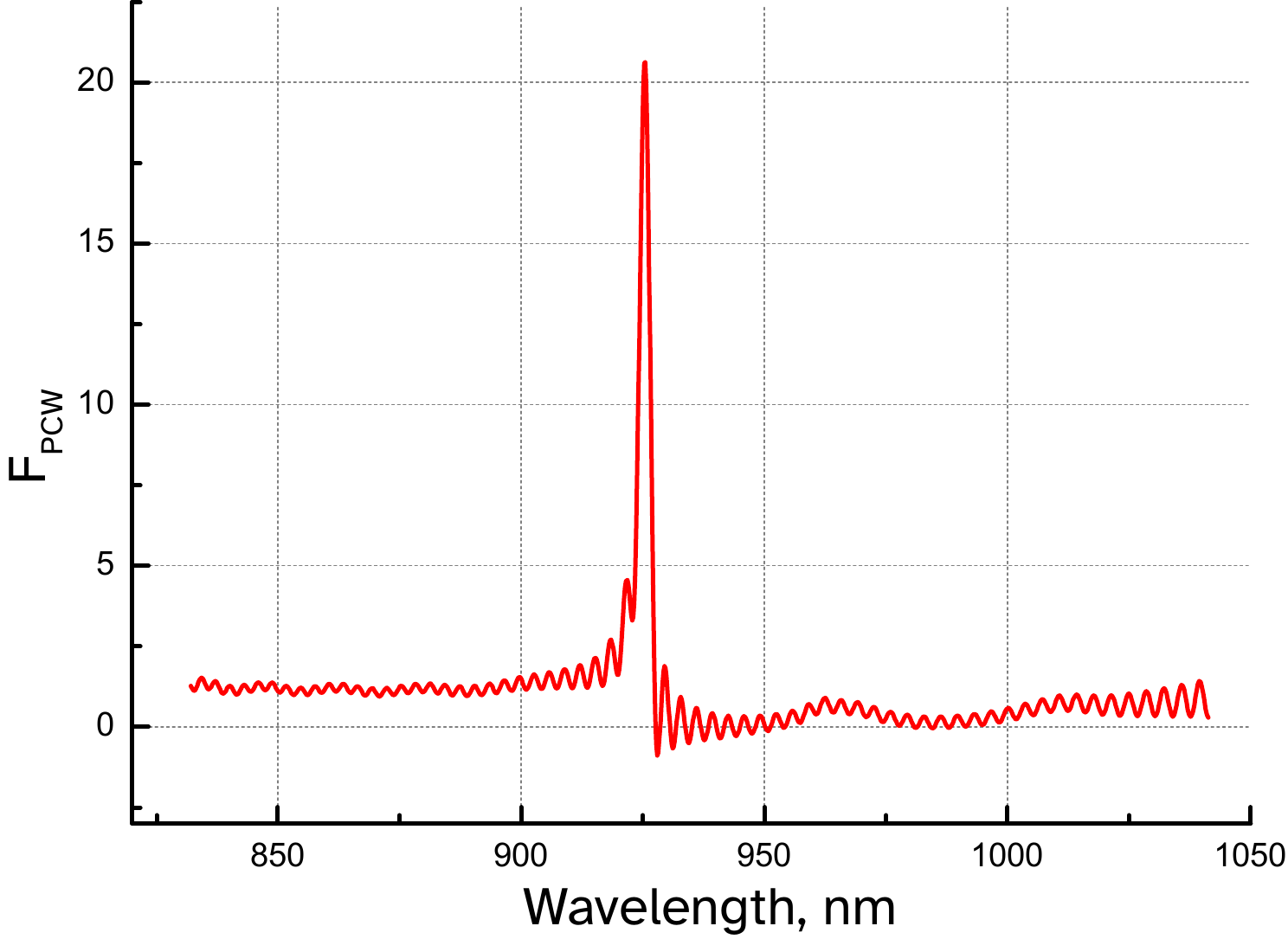}
 	\caption{The example of the numerically computed Purcell factor spectral dependece for a given crystal configuration.}
 	\label{fig:example_purcell_spectral}
 \end{figure}

We used a numerical model of a PC waveguide to compare to the predictions of our heuristic model. We set up the simulation of a PCW structure in Lumerical FDTD. We set the transverse dimensions of a GaAs membrane to be larger than the simulation region to avoid the internal reflections inside the crystal. We use PML boundary conditions along the waveguide axis as well as above and below the membrane. We do not set any specific Dirichlet boundary conditions \cite{Lodahl}, because we are not focused on precise calculation of the field distribution in the PCW structure. The PML boundary conditions must not overlay the features of the PCW in order to exclude numerical artifacts.
 
The PC waveguide dispersion was calculated in the following way. The simulation uses Bloch boundary conditions in the direction perpendicular to the waveguide axis. We use a dipole cloud as a source of radiation which excites multiple possible modes in the system. The randomly positioned electric field monitors record the local field amplitude versus the simulation time. The recorded traces are then decomposed as an infinite sum of exponentially decaying harmonic functions. Frequencies with the least decay rate are considered to be eigenfrequencies corresponding to a given wavevector modulus $k$, which is used to specify Bloch boundary conditions. 
 
The Purcell factor $F_{PCW}$ at the wavelength of interest was estimated using the built-in function of the Lumerical FDTD Software package. The example of the Purcel factor spectral dependence is illustrated in Fig.~\ref{fig:example_purcell_spectral}.

\end{document}